\documentclass[epjST]{svjour}
\usepackage{graphicx}
\usepackage{amsfonts,amssymb,amsmath}

\begin{document}
\title{Event coincidence analysis for quantifying statistical interrelationships between event time series}
\subtitle{On the role of flood events as triggers of epidemic outbreaks}
\author{Jonathan F. Donges\inst{1,2}\fnmsep\thanks{\email{donges@pik-potsdam.de}} \and Carl-Friedrich Schleussner\inst{1,3} \and Jonatan F. Siegmund\inst{1,4} \and Reik V. Donner\inst{1}}
\institute{Potsdam Institute for Climate Impact Research, Telegrafenberg A31, D-14473 Potsdam, Germany \and Stockholm Resilience Centre, Stockholm University, Kr\"aftriket 2B, 114 19 Stockholm, Sweden \and Climate Analytics, Friedrichstr. 231, Haus B, D-10969 Berlin, Germany \and Institute of Earth and Environmental Science, University of Potsdam, Karl-Liebknecht-Str. 24-25, D-14476 Potsdam-Golm, Germany}
\abstract{
Studying event time series is a powerful approach for analyzing the dynamics of complex dynamical systems in many fields of science. In this paper, we describe the method of event coincidence analysis to provide a framework for quantifying the strength, directionality and time lag of statistical interrelationships between event series. Event coincidence analysis allows to formulate and test null hypotheses on the origin of the observed interrelationships including tests based on Poisson processes or, more generally, stochastic point processes with a prescribed inter-event time distribution and other higher-order properties. Applying the framework to country-level observational data yields evidence that flood events have acted as triggers of epidemic outbreaks globally since the 1950s. Facing projected future changes in the statistics of climatic extreme events, statistical techniques such as event coincidence analysis will be relevant for investigating the impacts of anthropogenic climate change on human societies and ecosystems worldwide. (Date: \today)
} 
\maketitle
%

%
%
\section{Introduction}
\label{sec:intro}

Climate extremes and related natural disasters are of major interest for research on climate change and its impacts, because their frequency and amplitude is projected to increase significantly in the future~\cite{rahmstorf2011increase,coumou2012decade,IPCC2013}. However, when it comes to the quantification of impacts of associated natural disasters on ecosystems~\cite{rammig2015coincidences} and society (e.g. in terms of triggering epidemics or social unrest~\cite{Schleussner2015}), there are only very few studies providing a systematic assessment beyond individual cases. In-depth studies in this field require tailored  statistical analysis tools that allow for a quantitative characterization of statistical interdependencies between event time series and are also applicable to series comprising only a few events.

Time series of events or event series, here defined as an ordered set of $N$ event timings $\{t_1,\dots,t_N\}$, are the subject of study in many fields of science. In this paper, such event series are considered as binary, i.e. amplitudes associated to the event timings $t_i$ are either not available or are not taken into account in the analysis (corresponding to a description as unmarked point processes). There are many real-world examples of event series of this type, including photon arrival times in physics~\cite{Kuhn2002}, neuronal spikes in neurosciences~\cite{Quiroga2002,Brown2004}, exchange of messages on communication networks in social science~\cite{Wu2010} or timings of climatic extreme events~\cite{rahmstorf2011increase,coumou2012decade} and armed civil conflicts~\cite{hsiang2011civil,hsiang2013quantifying,hsiang2014climate} in climate impact studies~\cite{schellnhuber2013turn}. Many recent studies have focused on investigating statistical properties of single event series or point processes such as inter-event time distributions. For example, the analysis of human online communication reveals that waiting times between text messages do not follow an exponential distribution as expected from an uncorrelated (Poissonian) random process, but include bursts of frequent events interrupted by long periods of inactivity that can be better described by power-law distributions~\cite{Wu2010}.

However, less work appears to be available in the literature on quantifying and systematically studying statistical interrelationships between two or more event series, particularly when compared to the wide range of methods of this type available for standard time series such as Pearson correlation~\cite{BrockwellDavies2002}, mutual information~\cite{Kantz1997} or synchronization measures~\cite{Pikovsky2001}. Particularly in neuroscience, techniques have been developed for measuring the similarity or synchrony of event series of neuronal spike trains~\cite{Quiroga2002,Brown2004,Kreuz2007,Kreuz2007measuring,Dauwels2009}. In climatology, measures of event synchronization have been recently applied to study statistical interrelationships between extreme precipitation events and their complex spatial structure~\cite{Malik2011} using climate network approaches~\cite{Donges2015}. More specifically, this approach has been used to unravel the complex spatio-temporal patterns of heavy rainfall events in the Indian monsoon domain~\cite{Malik2011,Stolbova2014}, derive predictors for extreme flood events in South America~\cite{Boers2014prediction} and study regional climatological phenomena related to extreme precipitation over Central Europe~\cite{rheinwalt2015non}. 

Measures of event synchronization tend to be used mostly in an explorative mode of research aiming to reveal associations in large data sets of event series from neuroscience or climatology. However, some of the currently most debated problems in climate impact research, e.g. concerning climate-related variables such as extreme temperatures or the El Ni\~no--Southern Oscillation as potential drivers of armed civil conflicts~\cite{hsiang2011civil,hsiang2013quantifying}, call for a more in-depth analysis of statistical interrelations between event series. Extending upon previously applied event synchronization approaches, in this paper we formally put forward the alternative framework of event coincidence analysis~\cite{Donges2011c} for investigating in detail the statistical interrelationships between pairs of event time series and testing hypotheses on the nature of these interrelationships. Event coincidence analysis is designed to measure the strength, directionality and time lag of statistical relations between event series. The method was introduced in a less general setting to study possible statistical interrelationships between nonlinear regime shifts in African paleoclimate during the past 5\,million years and events in hominin evolution such as the appearance and disappearance of species~\cite{Donges2011c}. It has also been applied to investigate the impacts of climatic extremes such as droughts and heat waves on vegetation productivity based on observational data and dynamic vegetation model runs~\cite{rammig2015coincidences}. Furthermore, event coincidence analysis has been used to evaluate different hypotheses on socio-economic factors influencing the vulnerability of countries to natural disasters with a focus on the possible triggering of outbreaks of civil conflicts~\cite{Schleussner2015}.

We argue that event coincidence analysis is a particularly useful tool in the area of climate impact studies, since it allows to statistically study the effects of such extreme events on other processes and explicitly takes their nature as event series into account. To illustrate the capabilities of our approach, we employ event coincidence analysis to assess extreme flood events as possible drivers of epidemics extending upon earlier work~\cite{rammig2015coincidences,Schleussner2015}. Applying the framework in this case study based on observational data yields evidence that, from a globally aggregated perspective, flood events have acted as drivers of epidemics in the same country in the past.

The structure of this paper is as follows: event coincidence analysis is thoroughly introduced in Sect.~\ref{sec:methods} including descriptions of the basic methodology, statistical null models for testing hypotheses and related approaches. Subsequently, the results of applying event coincidence analysis to event series of extreme floods and epidemics are reported in Sect.~\ref{sec:applications}. Finally, Sect.~\ref{sec:conclusions} provides conclusions and perspectives for promising future extensions of the event coincidence analysis methodology.

%
%
\section{Methods}
\label{sec:methods}

\begin{figure}[tb]
\centering
\includegraphics[width=0.75\textwidth]{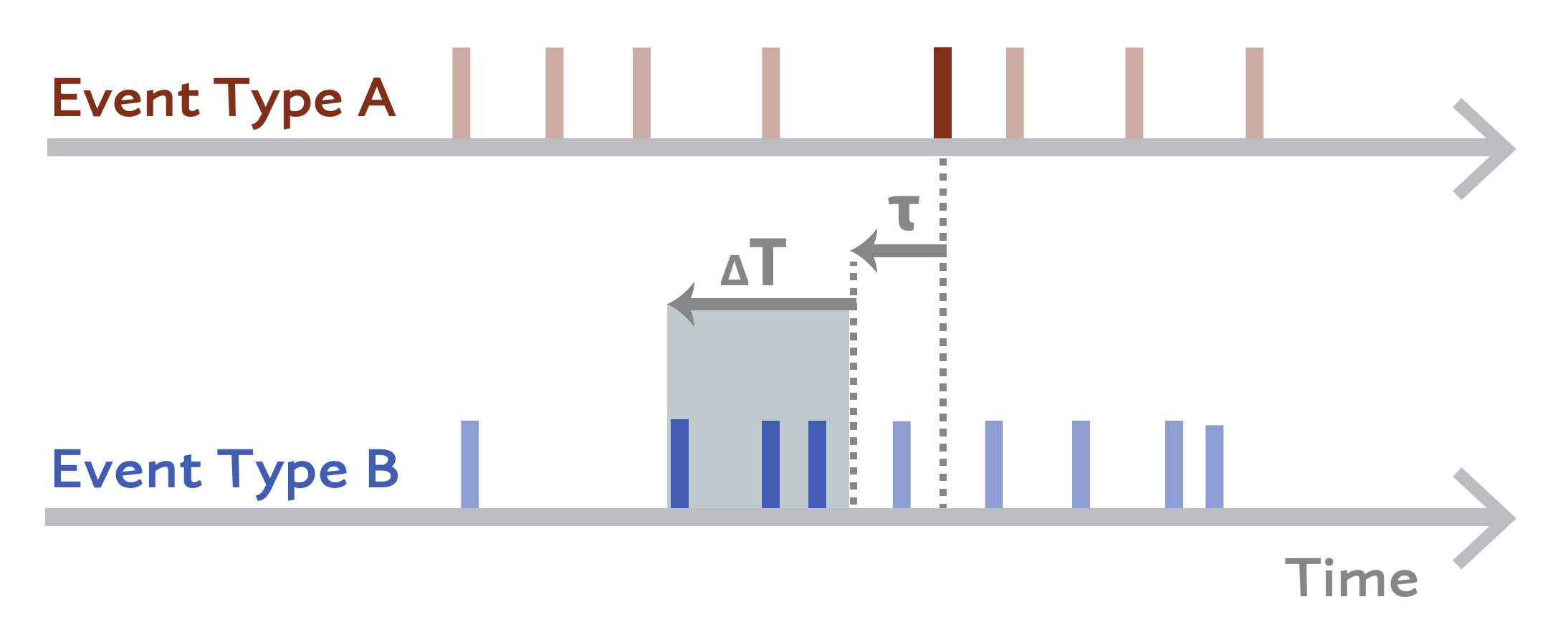} 
\caption{Schematic illustration of event coincidence analysis for quantifying statistical interrelationships between two event time series $A$ and $B$ for the case of precursor coincidences. The assumption to be quantified and tested for is that events in $B$ are precursors of events in $A$ (under the condition that an $A$-event has occurred). Focusing on an event in series $A$ (dark red bar), a (lagged) coincidence occurs with events in series $B$ (dark blue bars) if the latter fall into the coincidence interval of width $\Delta T$ (grey bar) that can be shifted by a lag parameter $\tau$. Coincidence rates are obtained by computing the relative frequency of occurrence of such coincidences for all events in series $A$ (Sect.~\ref{sec:methods}).}
\label{fig:coincidence_scheme}
\end{figure}

In this section, we develop the method of event coincidence analysis that is concerned with quantifying the statistical interrelationships between pairs of event series, extending upon the approach introduced in~\cite{Donges2011c}. A pair of \emph{event time series} $A$ and $B$ is here defined as two ordered event sets with timings $\{t_1^A,\dots,t_{N_A}^A\}$ and $\{t_1^B,\dots,t_{N_B}^B\}$ with numbers of events $N_A$, $N_B$, respectively. Both event series are assumed to cover a time interval $(t_0, t_f)$ of length $T = t_f-t_0$, such that $t_0 \leq t_1^A \leq \dots \leq t_{N_A}^A \leq t_f$ and $t_0 \leq t_1^B \leq \dots \leq t_{N_B}^B \leq t_f$. This yields event rates $\lambda_A = N_A / T$ and $\lambda_B = N_B / T$. 

Event coincidence analysis is based on counting coincidences between events of different types. In the following, the assumption to be quantified and tested for is that events in $B$ precede events in $A$, which is related to a possible causal influence from $B$- to $A$-type events  (Fig.~\ref{fig:coincidence_scheme}). The opposite case of assuming that events in $A$ precede events in $B$ can be accommodated by exchanging the labels $A$ and $B$ throughout the formulae and text. 

An \emph{instantaneous coincidence} is defined to occur if two events at $t_i^A, t_j^B$ with $t_j^B < t_i^A$ are closer in time than a temporal tolerance or \emph{coincidence interval} $\Delta T$, i.e. if 
\begin{align}
t_i^A - t_j^B \leq \Delta T \label{eq:coincidence}
\end{align}
holds. In turn, a \emph{lagged coincidence} is defined as an instantaneous coincidence between the time shifted event at $t_i^A - \tau$, where $\tau \geq 0$ is a time lag parameter, and the event at $t_j^B < t_i^A - \tau $, i.e. if the condition 
\begin{align}
(t_i^A - \tau) - t_j^B \leq \Delta T \label{eq:lagged_coincidence}
\end{align}
is satisfied. 

Differing from the above problem formulation that is consistently used throughout this work, we note that event coincidence analysis can also be performed by employing coincidence intervals that are symmetric around $A$-events and relaxing the assumption that $B$-events must precede $A$-events. The resulting condition $|t_i^A - t_j^B| \leq \Delta T $ can be meaningful, e.g. given event series with pronounced dating uncertainties as in the case of archeological, paleontological and paleoenvironmental data~\cite{Donges2011c}.

In the following, we introduce the concept of the \emph{coincidence rate} between a single pair of event series (Sect.~\ref{sec:ca}) as well as an aggregated coincidence rate for taking into account several pairs of event series (Sect.~\ref{sec:agg_ca}). Section~\ref{sec:null_models} discusses coincidence statistics for null models of stochastic point processes that can be used to test the statistical significance of coincidence rates estimated from data. Moreover, we put event coincidence analysis into the context of other related approaches for the analysis of event time series (Sect.~\ref{sec:related_methods}). 

Many of the measures and significance tests described below are implemented in the open source software package \texttt{CoinCalc}~\cite{Siegmund2015b} written in the programming language R which is available at \texttt{https://github.com/JonatanSiegmund/CoinCalc}.

\subsection{Coincidence rates for a pair of event series}
\label{sec:ca}

For quantifying the strength of statistical interrelationships between two event time series $A$ and $B$, we introduce two variants of coincidence rates addressing $B$-type events as \emph{precursors} and \emph{triggers} of $A$-type events, respectively. In the first case, the \emph{precursor coincidence rate}
\begin{align}
r_p(\Delta T, \tau) = \frac{1}{N_A} \sum_{i=1}^{N_A} \Theta \left[ \sum_{j=1}^{N_B} 1_{[0,\Delta T]}\left((t_i^A - \tau) - t_j^B\right) \right],
\end{align}
measures the fraction of $A$-type events that are preceded by at least one $B$-type event (note that multiple $B$-type events within the coincidence interval are counted only once, see also Fig.~\ref{fig:coincidence_scheme}). Here, $\Theta(\cdot)$ denotes the Heaviside function (here defined as $\Theta(x)=0$ for $x\leq0$ and $\Theta(x)=1$ otherwise) and $1_I(\cdot)$ the indicator function of the interval $I$ (defined as $1_I(x)=1$ for $x \in I$ and $1_I(x)=0$ otherwise). In the second case, the \emph{trigger coincidence rate}
\begin{align}
r_t(\Delta T, \tau) = \frac{1}{N_B} \sum_{j=1}^{N_B} \Theta \left[ \sum_{i=1}^{N_A} 1_{[0,\Delta T]}\left((t_i^A - \tau) - t_j^B\right) \right],
\end{align}
measures the fraction of $B$-type events that are followed by at least one $A$-type event (note that multiple $A$-type events within the coincidence interval are counted only once). Distinguishing between precursor and trigger coincidence rates allows to introduce a certain notion of directionality to the method of event coincidence analysis. Furthermore, the parameter $\tau$ allows to explicitly take into account lagged relationships between event series.

\subsection{Aggregated coincidence rates}
\label{sec:agg_ca}

In some applications, it can be relevant to have at hand an integrated measure for coincidences that occur between several pairs of event series as in the case of events that are available on a spatial grid or for different regions or countries. For example, consider multiple country-wise sets of $A$-type events (floods) and $B$-type events (epidemic outbreaks) as in the application presented in Sect.~\ref{sec:applications}. In this case, coincidences can only be meaningfully counted on a per-country basis, but it is desirable to quantify the aggregated coincidence rate over all countries in the data set or a suitably filtered subset of countries to obtain a global measure of the strength of the relationship between the two event types considered and its statistical significance with respect to different null hypotheses~\cite{Schleussner2015}. Another motivation for considering aggregate measures of coincidence relationships is related to data quality. In some applications with small event numbers $N_A, N_B$, only aggregation over several pairs of event series allows to draw robust statistical conclusions.

Analogously to the case of a single pair of event series, two flavors of aggregated coincidence rates are defined as follows given a set $G$ of pairs of $A$- and $B$-type events. The \emph{aggregated precursor coincidence rate}
\begin{equation}
r_p^G(\Delta T, \tau) = \frac{\sum_{k \in G} \sum_{i=1}^{N_{A,k}} \Theta \left[ \sum_{j=1}^{N_{B,k}} 1_{[0,\Delta T]}\left((t_i^{A,k} - \tau) - t_j^{B,k}\right)\right]}{\sum_{k \in G} N_{A,k}}
\end{equation}
measures the total number of precursor coincidences occurring in all pairs of event series normalized by the maximum possible number of such coincidences. Along the same lines, the \emph{aggregated trigger coincidence rate}
\begin{equation}
r_t^G(\Delta T, \tau) = \frac{\sum_{k \in G} \sum_{i=1}^{N_{B,k}} \Theta \left[ \sum_{j=1}^{N_{A,k}} 1_{[0,\Delta T]}\left((t_i^{A,k} - \tau) - t_j^{B,k}\right) \right]}{\sum_{k \in G} N_{B,k}}
\end{equation}
is the accordingly normalized total number of trigger coincidences occurring in all pairs of event series in $G$. Note that for both types of aggregated coincidence rates, multiple events falling within the coincidence window are counted only once, as in the definition of coincidence rates for a single pair of event series (Sect.~\ref{sec:ca}).

Studying aggregated coincidence rates can be seen as a first step towards a systematic analysis of coincidences in more general spatio-temporal event data. Such data can be conceptionalized as being generated by spatial or spatio-temporal point processes~\cite{Moller2003,diggle2006spatio}. More generally, events of interest for an extended event coincidence analysis can also take the form of higher dimensional objects with a nontrivial shape in terms of, e.g. latitude, longitude and time, such as the spatio-temporal extremes in the fraction of absorbed photosynthetically active radiation (fAPAR) identified by Zscheischler et al.~\cite{zscheischler2013detection}.

\subsection{Statistics for null models of stochastic point processes}
\label{sec:null_models}

Treating stochastic point processes as generators of event time series allows to derive distributions of coincidence rates to test the statistical significance of the results of event coincidence analysis based on a hierarchy of null hypotheses, analogously to classical statistics and the method of surrogates for standard time series analysis~\cite{Kantz1997,Schreiber2000}. Here, we focus on Poisson processes without temporal correlations between events (Sect.~\ref{sec:poisson}) and point processes with a prescribed inter-event time distribution $P(\Delta t)$ that allow to consider, e.g. processes with heavy-tailed $P(\Delta t)$ that tend to produce bursts of events (Sect.~\ref{sec:short_term_correlated}). More generally, classes of null models for point processes of interest include, for example, event series with higher-order memory effects such as correlations between event bursts~\cite{jo2015correlated} that are, however, beyond the scope of this paper. We also briefly touch upon the possibility of constructing surrogate event series from time series surrogates (Sect.~\ref{sec:event_series_from_ts_surrogates}). 

Analytical results are given below where available, otherwise we rely on Monte Carlo simulations. For illustration, we restrict ourselves to a single pair of event time series, the case of sets of event series can be treated analogously. Significance tests based on the null hypothesis of Poisson processes following a Monte Carlo approach are applied in Sect.~\ref{sec:applications} to quantify the statistical interrelationships between flood events and epidemic outbreaks.

\subsubsection{Poisson processes}
\label{sec:poisson}

Here, we assume that both $A$- and $B$-type events are generated by Poisson processes with event rates $\lambda_A$ and $\lambda_B$, respectively. This assumption implies that both types of events are distributed randomly, independently and uniformly over the continuous time interval of length $T$. Since our focus is on using the derived statistics for hypothesis testing on data sets with typically small numbers of events in each series, we assume fixed event numbers $N_A = \lambda_A T$ and $N_B = \lambda_B T$. Note that the analytically derived estimators presented below are only expected to yield reliable results in the limit of sufficiently large event numbers 
\begin{align}
N_A \gg 1 \text{ and } N_B \gg 1. \label{eq:cond_event_numbers}
\end{align}

First, we analytically derive the statistics of precursor coincidence rates extending upon~\cite{Donges2011c}. The probability for a (lagged) precursor coincidence between an $A$-event and a preceding $B$-event is given by the probability
\begin{align}
p = \frac{\Delta T}{T-\tau}
\end{align}
that a $B$-event occurs randomly in a segment of length $\Delta T$ of the effective time span of interest $T-\tau$. This follows from the null hypothesis of Poisson processes generating the event series, where the probability for events to occur is the same in any time instant and is independent from the occurrence of other events, resulting in a linear dependence of $p$ on $\Delta T$.

Then the probability of a specific $A$-event to coincide with \emph{at least one} of the $N_B$ $B$-events is given by
\begin{align}
1 - (1-p)^{N_B} = 1 - \left(1-\frac{\Delta T}{T-\tau}\right)^{N_B}.
\end{align}
Note that when counting only exactly contemporaneous coincidences with $\Delta T = 0$, $p=0$ follows in the limit of a continuous time axis. However, in real-world data sets, the time axis is often discrete, e.g. due to finite sampling intervals or finite numerical precision. In this common case, $p = 1 / (T-\tau)$ needs to be used in the following with $T$ and $\tau$ measured in numbers of time steps instead of units of absolute time when the interest is in coincidences with zero tolerance~\cite{Siegmund2015}.

Based on this expression, we can calculate the probability $P(K; N_A, 1 - (1-p)^{N_B})$ that exactly $K$ precursor coincidences are observed for a given realization of the two Poisson processes. Even though $A$-events are assumed to be distributed independently in the interval $\left[0,T\right]$, to proceed with the derivation we need to further assume that $A$-type events are typically spaced much more widely than the coincidence interval $\Delta T$, i.e. 
\begin{align}
\Delta T \ll T / N_A. \label{eq:poisson_process_conditions_precursor}
\end{align}
When this condition (Eq.~\ref{eq:poisson_process_conditions_precursor}) is fulfilled, the events that a specific $A_i$-event coincides with \emph{at least one} $B$-event and that another specific $A_j$-event coincides with \emph{at least one} $B$-event can be considered statistically independent. Only then, $P(K; N_A, 1 - (1-p)^{N_B})$ is given by the binomial distribution with $N_A$ trials and a success probability $1 - (1-p)^{N_B}$~\cite{StatisticsBook} and, hence,
\begin{align}
P(K; N_A, 1 - (1-p)^{N_B}) = \binom{N_A}{K} \left(1 - \left(1-\frac{\Delta T}{T-\tau}\right)^{N_B}\right)^K \left(\left(1-\frac{\Delta T}{T-\tau}\right)^{N_B}\right)^{N_A-K}. \label{eq:binom_dist}
\end{align}
Using the relationship $K=r_p N_A$ to substitute $K$ by $r_p$ in the above equation yields the distribution of precursor coincidence rates $P(r_p; N_A, 1 - (1-p)^{N_B})$.

From the distribution (Eq.~\ref{eq:binom_dist}), the expectation value $\left<K\right>$ and standard deviation $\sigma(K)$ can be straightforwardly derived as
\begin{align}
\left<K\right> = N_A \left(1 - (1-p)^{N_B}\right) = N_A \left(1- \left(1-\frac{\Delta T}{T-\tau}\right)^{N_B}\right)
\end{align}
and 
\begin{align}
\sigma(K) &= \sqrt{N_A \left(1 - \left(1-p\right)^{N_B}\right) \left(1-p\right)^{N_B}} \nonumber \\
&= \sqrt{N_A \left(1 - \left(1-\frac{\Delta T}{T-\tau}\right)^{N_B}\right) \left(1-\frac{\Delta T}{T-\tau}\right)^{N_B}}.
\end{align}
This yields the expectation value of the precursor coincidence rate
\begin{align}
\left<r_p\right> = \frac{\left<K\right>}{N_A} = 1- \left(1-\frac{\Delta T}{T-\tau}\right)^{N_B} \label{eq:prec_coincidence_rate}
\end{align}
and its standard deviation
\begin{align}
\sigma(r_p) &= \sigma(K) / N_A \nonumber \\
&= \sqrt{\frac{1}{N_A} \left(1 - \left(1-\frac{\Delta T}{T-\tau}\right)^{N_B}\right) \left(1-\frac{\Delta T}{T-\tau}\right)^{N_B}}. \label{eq:std_prec_coincidence_rate}
\end{align}

The $p$-value of an observation $K_e$ with respect to the test distribution (Eq.~\ref{eq:binom_dist}), i.e. the probability to obtain a number of coincidences $K$ larger or equal to the empirically observed number $K_e$, is then given by
\begin{align}
P(K \geq K_e) = \sum_{K^\ast=K_e}^{N_A} P(K^\ast; N_A, 1-(1-p)^{N_B}).
\end{align}

The statistics of trigger coincidence rates for event series generated by Poisson processes can be derived analogously by assuming a wide enough typical spacing of $B$-events:
\begin{align}
\Delta T \ll T / N_B. \label{eq:poisson_process_conditions_trigger}
\end{align}
The distribution of the number of trigger coincidences $K$ is then given by
\begin{align}
P(K; N_B, 1 - (1-p)^{N_A}) = \binom{N_B}{K} \left(1 - \left(1-\frac{\Delta T}{T-\tau}\right)^{N_A}\right)^K \left(\left(1-\frac{\Delta T}{T-\tau}\right)^{N_A}\right)^{N_B-K}
\end{align}
yielding the expectation value and standard deviation of the trigger coincidence rate
\begin{align}
\left<r_t\right> = 1- \left(1-\frac{\Delta T}{T-\tau}\right)^{N_A}
\end{align}
and
\begin{align}
\sigma(r_t) = \sqrt{\frac{1}{N_B} \left(1 - \left(1- \frac{\Delta T}{T-\tau}\right)^{N_A}\right) \left(1-\frac{\Delta T}{T-\tau}\right)^{N_A}},
\end{align}
respectively. As above, the $p$-value of an empirically observed number of trigger coincidences $K_e$ can then be written as
\begin{align}
P(K \geq K_e) = \sum_{K^\ast=K_e}^{N_B} P(K^\ast; N_B, 1-(1-p)^{N_A}).
\end{align}

\begin{figure}[tb]
\centering
\includegraphics[width=0.49\textwidth]{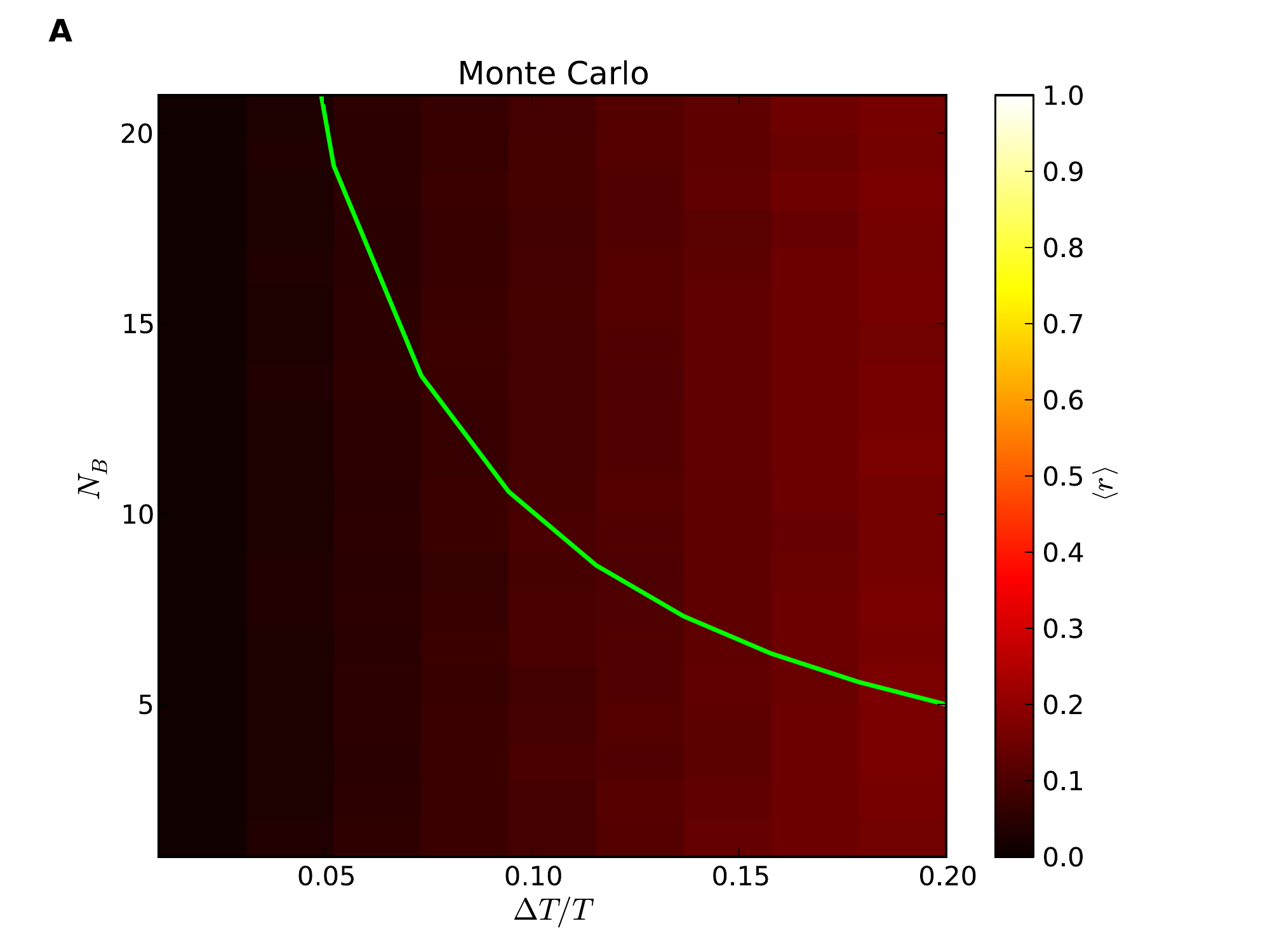}
\includegraphics[width=0.49\textwidth]{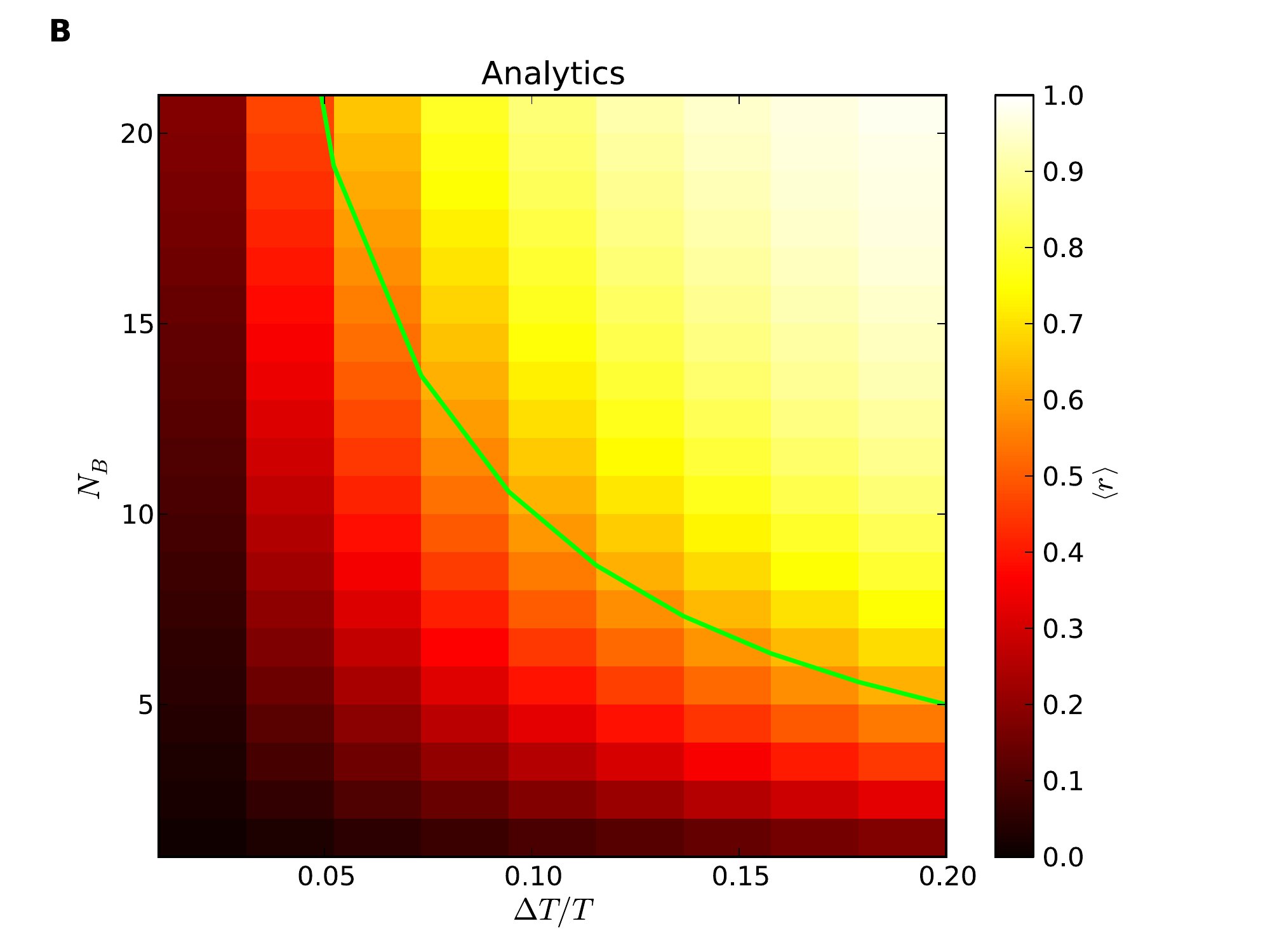} \\
\includegraphics[width=0.49\textwidth]{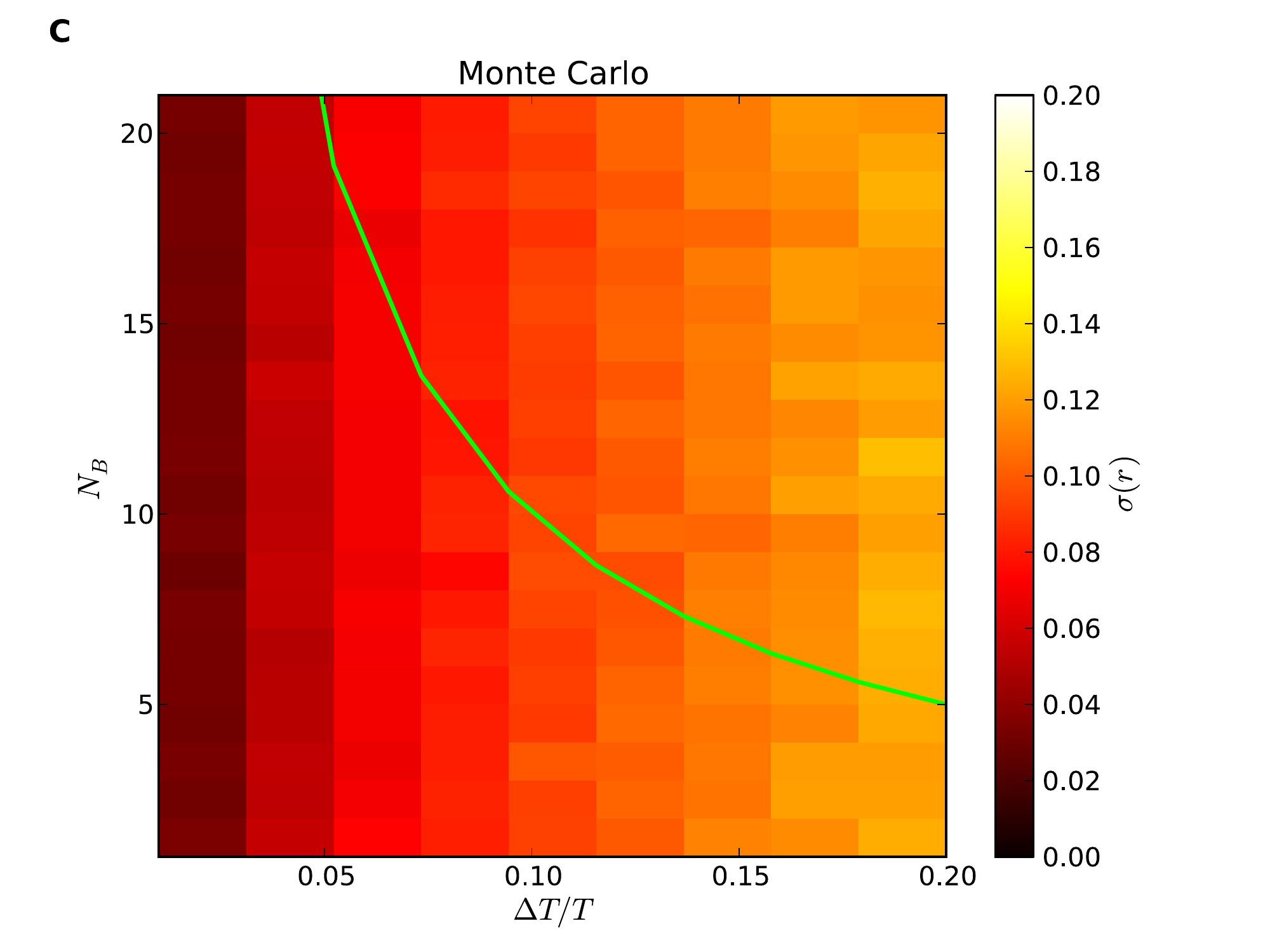}
\includegraphics[width=0.49\textwidth]{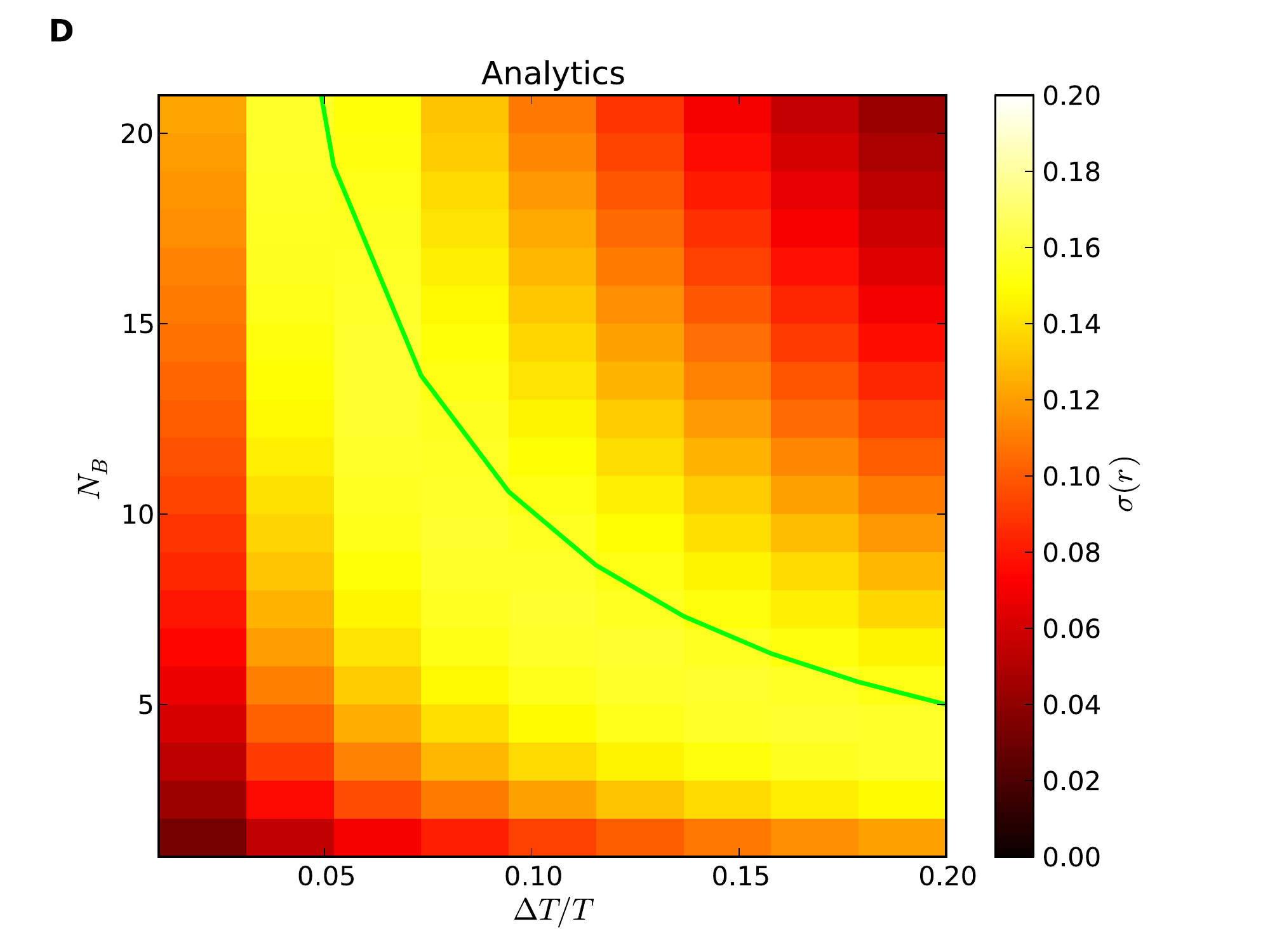}
\caption{Comparison of the expected trigger coincidence rate~$\left<r_p\right>$ (A,B) and its standard deviation $\sigma(r_p)$ (C,D) obtained using Monte Carlo simulation (A,C) and an analytical approximation (B,D) depending on the relative coincidence interval $\Delta T / T$ and the number of $B$-type events $N_B$. The analytical approximation is only accurate in the regime $N_B \gg 1$ and $N_B \ll N_B^c (\Delta T / T) = (\Delta T / T)^{-1}$ (green line), where $N_B^c(\cdot)$ denotes a critical value of $N_B$. In this example, the number of $A$-type events is $N_A=10$, no lag is used ($\tau=0$), events are distributed in the unit interval of width $T=1$ and $m=1,000$ trials are used in the Monte Carlo simulations for each considered combination of parameters.}
\label{fig:mc_analytics_comparison}
\end{figure}

In the case that the conditions (\ref{eq:cond_event_numbers}), (\ref{eq:poisson_process_conditions_precursor}) and (\ref{eq:poisson_process_conditions_trigger}) are not met, Monte Carlo simulations need to be applied to compute statistics of event coincidence analysis such as the mean and standard deviation of coincidence rates or the significance level ($p$-value) of an observed coincidence rate corresponding to the null hypothesis that the empirical coincidence rate can be explained as the result of Poisson processes. To illustrate this issue, we compare the expectation values and standard deviations of trigger coincidence rates for Poisson processes derived from analytics and Monte Carlo simulation for different relative coincidence intervals $\Delta T / T$ and numbers of $B$-events $N_B$ (Fig.~\ref{fig:mc_analytics_comparison}). Indeed, the statistics are only comparable if conditions (\ref{eq:cond_event_numbers}) and (\ref{eq:poisson_process_conditions_trigger}) are met, i.e. for $N_B \gg 1$ and $N_B \ll N_B^c (\Delta T / T) = (\Delta T / T)^{-1}$ (green line in Fig.~\ref{fig:mc_analytics_comparison}), where $N_B^c(\cdot)$ denotes a critical value of $N_B$. Otherwise, Monte Carlo simulations show that the analytical approximation tends to overestimate the expected coincidence rate and its standard deviation, even though the approximations (Eqs.~\ref{eq:prec_coincidence_rate}, \ref{eq:std_prec_coincidence_rate}) show the correct asymptotic behavior of $r \to 0$ and $\sigma(r) \to 0$ for $\Delta T / T \to 0$ and $r \to 1$ and $\sigma(r) \to 0$ for $\Delta T / T \to 1$.

\subsubsection{Stochastic point processes with prescribed inter-event time distribution}

\label{sec:short_term_correlated}

Compared to the Poisson processes discussed above, a more general null hypothesis is that the observed values of coincidence rates can be explained by stochastic point processes with a given distribution of inter-event times $P(\Delta t)$. For example, the inter-event time distribution for Poisson processes with average event rate $\lambda$ is given by the exponential distribution
\begin{align}
P_1(\Delta t) = \lambda e^{-\lambda \Delta t}.
\end{align}
However, it has been shown that many event time series display bursting behavior associated with inter-event time distributions having more slowly decaying (heavy) tails than the exponential distribution $P_1(\Delta t)$ \cite{Wu2010,Picoli2014}. For example, human violent conflicts were reported to display universal bursting behavior~\cite{Picoli2014} associated with inter-event time distributions of the form
\begin{align}
P_2(\Delta t) = \lambda F(\lambda \Delta t),
\end{align}
where $F(\cdot)$ exhibits a power-law decay with exponent $\alpha$ such that
\begin{align}
F(x) = a x^{-\alpha}
\end{align}
yielding
\begin{align}
P_2(\Delta t) \propto \Delta t^{-\alpha}.
\end{align}
Power-law inter-event time distributions with an exponential cutoff 
\begin{align}
P_3(\Delta t) = C (\lambda \Delta t)^\alpha e^{-\lambda \Delta t/\beta}
\end{align}
have been reported to accurately describe the return time statistics of earthquakes \cite{corral2004long} and other types of event series.

While deriving analytical results for coincidence statistics based on these and other classes of point processes with prescribed $P(\Delta t)$ remains the subject of future research, Monte Carlo simulations can be applied to obtain test distributions for assessing the statistical significance of empirically observed coincidence rates. Note that these tests can only be meaningfully applied in practice if either $P(\Delta t)$ can be estimated well from the empirically observed inter-event time statistics requiring a sufficiently large number of events, or a good process understanding exists, i.e. the inter-event time distribution is known from theoretical considerations or observations from analogous systems. Alternatively, ensembles of surrogate event series can be generated by randomly shuffling inter-event time intervals given a large number of events. These conditions are not met for the application studied in Sect.~\ref{sec:applications}, implying the need for restricting the analysis to the null hypothesis of Poisson processes there.

\subsubsection{Surrogate event series generated from time series surrogates}

\label{sec:event_series_from_ts_surrogates}

In a number of relevant applications of event coincidence analysis, e.g. when studying climatological extreme events, event series are generated from underlying time series data. This transformation from time series to event data is typically achieved by thresholding to identify extreme events in the time series according to a prescribed quantile~\cite{rammig2015coincidences,Siegmund2015} or some other form of filtering. In this case, various types of time series surrogates~\cite{Schreiber2000} can be used to generate ensembles of event series for hypothesis testing by applying the same transformation to original and surrogate time series data. For example, univariate iterative amplitude adjusted Fourier transform (iAAFT) surrogates as implemented in~\cite{donges2015unified} can be used to generate surrogate event series based on surrogate time series with the same amplitude distribution and autocorrelation function as the original data. This procedure is useful for constructing suitable significance tests for event coincidence analysis when extreme events in the series of interest tend to cluster due to pronounced autocorrelation in the underlying time series data, as it was found to be the case for European temperature, precipitation, tree ring width and simulated net primary productivity (NPP)~\cite{rammig2015coincidences}. Along these lines, bivariate event series surrogates derived from bivariate iAAFT time series surrogates could be used for testing the null hypothesis that observed coincidence rates can be explained by the co-occurrence of extremes due to the conserved linear cross-correlation structure of the underlying pair of time series.

\subsection{Related methods}
\label{sec:related_methods}

The complex systems-inspired framework of event coincidence analysis presented above is conceptually related to measures from spatial statistics for the correlation of spatial and spatio-temporal point processes~\cite{Moller2003,diggle2006spatio} such as Ripley's cross-$K$ \cite{Dixon2006} as well as various forms of regression analysis for point process data. Another popular related approach is considering measures of event synchronization for quantifying the similarity of event series~\cite{Kreuz2007,Malik2011,Boers2014prediction}. Hence, the considerations on surrogate event series and significance tests given above could be applied to the latter concept as well, given that the requirements and basic assumptions are met. However, it should be noted that event synchronization lacks the distinction between coincidence interval $\Delta T$ and lag parameter $\tau$ provided by event coincidence analysis, and also does not allow to distinguish the cases of precursor and trigger coincidences.

While the statistical theory of temporal point processes appears generally less consolidated than the theory of standard time series~\cite{Brown2004}, a multitude of methodologies for studying statistical interrelationships between event time series have been developed in the neurosciences in the last decades focussing on the specific, but important, application to neural spike trains. These techniques include methods focussing on the distributions of relative waiting times of events in series $A$ with respect to events in series $B$~\cite{Perkel1967}, cross-correlograms and cross-intensity functions as well as frequency-domain methods, neural spike train decoding or information-theoretical methods~\cite{Brown2004}. Certainly, this wealth of alternative methodologies holds a great potential for fruitful applications in other fields of science, considering, for example, event series of climatological extreme events and natural disasters.

It should also be mentioned that the statistical and mathematical literature contains a large number of less closely related studies of coincidences, e.g. considering the \emph{birthday problem}~\cite{Diaconis1989}. The term \emph{coincidence analysis} is also used in different contexts in fields such as elementary particle physics~\cite{zaborov2009coincidence} or in the identification of causal dependencies in configurational data~\cite{baumgartner2015identifying}. This is why we choose to use the more specific term \emph{event coincidence analysis} when referring to the methodology introduced in this paper.

\section{Application: extreme flood events as possible drivers of epidemics}
\label{sec:applications}

To illustrate the capabilities of event coincidence analysis, we apply it here to analyze the interrelations between two types of event time series of natural disasters, for which a causal relation is commonly assumed in the literature: hydrological flooding events ($B$-events) and outbreaks of epidemics ($A$-events) \cite{Campbell-Lendrum2014}. Our analysis is performed on the EmDAT data base covering the time interval 1950--2009 in a monthly time resolution \cite{emdat}. This data base contains 3,468 flood events worldwide that are defined as a significant rise of water level in a stream, lake, reservoir or coastal region as well as 1,152 epidemic outbreaks, defined as either an unusual increase in the number of cases of an infectious disease that already exists in the region or population concerned, or the appearance of an infectious disease previously absent from a region. For each country $k$ in the data base, a pair of event series is available containing $N_{f,k}$ flood events and $N_{e,k}$ epidemic outbreaks, respectively.

As described above, event coincidence analysis allows for two different test setups: in the first setup, we test on the basis of the occurrence of epidemic outbreaks and perform a coincidence test with flood events preceding epidemic outbreaks within a given coincidence interval (statistics based on precursor coincidences). Since we analyze coincidences based on the condition that an epidemic outbreak has occurred, this setup may also be termed a \emph{risk enhancement test}~\cite{Schleussner2015}. In the second case, we perform the event coincidence analysis on the basis of occurrence of flood events that are followed by epidemic outbreaks (statistics based on trigger coincidences). We call this the \emph{trigger test}~\cite{Schleussner2015}, since it investigates a possible causal direction of flood events triggering epidemic outbreaks. In the following, we do not consider additional time lags between different types of events that are not covered already by the coincidence interval $\Delta T$ and, hence, set $\tau=0$ (see Fig.~\ref{fig:coincidence_scheme}). Furthermore, we compute aggregated coincidence rates covering all countries in the data base (Sect.~\ref{sec:agg_ca}) as well as coincidence rates on a country-wise basis (Sect.~\ref{sec:ca}).

To test for statistical significance with the null hypothesis (NH) that the observed coincidences can be explained on the basis of event series generated by Poisson processes with the empirically observed event rates, Monte Carlo simulation is applied to generate pairs of surrogate event time series with conserved event numbers on an individual country basis by uniformly and independently drawing $N_{f,k}, N_{e,k}$, event timings over the full analysis period 1950--2009. We generate $m=1,000$ ensemble members for each country and significance levels of 95\,\% and 99\,\% are applied for the rejection of the NH.

\begin{figure}[tb]
	\centering
	\includegraphics[width=0.75\textwidth]{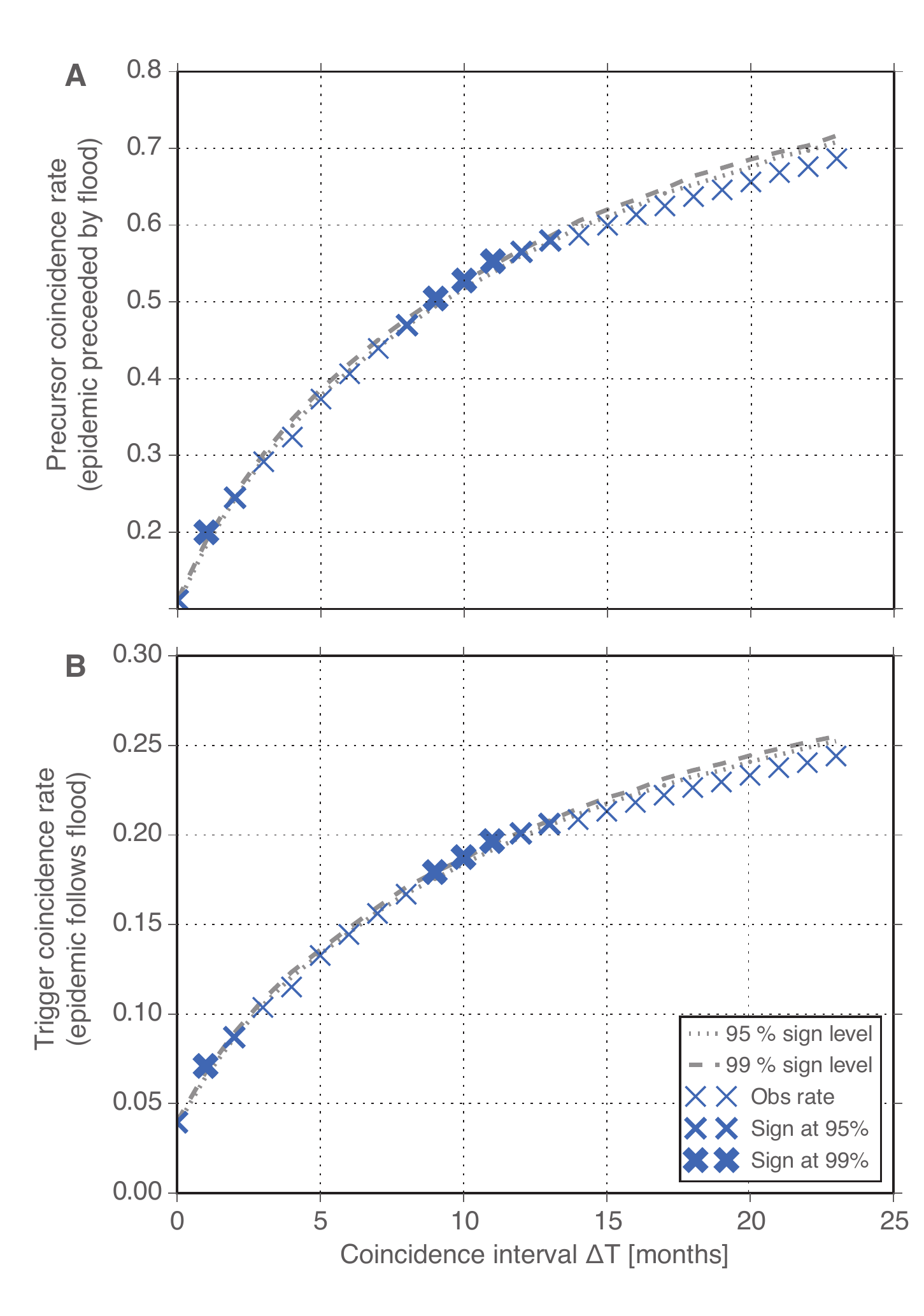}
	\caption{Results of event coincidence analysis for the flood and epidemic outbreak event time series: Aggregated precursor (A) and trigger (B) coincidence rates. Dotted (dashed) grey lines mark the 95\,\% (99\,\%) significance level determined by Monte Carlo simulations. Coincidence rates that are significant at 95\,\% (99\,\%) levels are highlighted by bold markers.}
	\label{fig:results_epidemics_test}
\end{figure}

Figure \ref{fig:results_epidemics_test} displays coincidence rates aggregated over all countries with available event data for coincidence intervals $\Delta T$ ranging from 0 to 24 months. While $\Delta T = 0$ implies considering coincidences within the same month, a 24-months window counts coincidences between 0 and 24 months after (before) a flooding (epidemic outbreak) event, respectively. For the risk enhancement test, we find that about 20\,\% of all epidemic outbreaks have been preceded by a flooding event within a month before the outbreak. Our corresponding results indicate that floods robustly contribute to the outbreak risk (Fig.~\ref{fig:results_epidemics_test}A). While no direct causal attribution is possible based on this test, the trigger test (Fig.~\ref{fig:results_epidemics_test}B) also robustly suggests a possible causal relationship. We find that about 7\,\% of all flooding events have been followed by an epidemic outbreak in the next month, which is significant at the 99\,\% level. 

These results are robust also for other small coincidence intervals. In turn, for window widths $\Delta T$ exceeding 3 months, we do not find indications that the NH of the aggregated coincidence rate arising by chance can be rejected. However, this changes for window lengths between 9 and 13 months, where the NH can be rejected at least at the 95\,\% level, indicating a robust long-range interrelation between the two types of events. In fact, more than 50\,\% of all epidemic outbreaks have been preceded by a flooding event over a 12-months coincidence interval and about 20\,\% of all flooding events have possibly triggered epidemic outbreaks within the 12 months following the natural disaster.

It should be noted that although we perform multiple hypothesis tests for varying coincidence intervals $\Delta T$, standard corrections of the significance level to account for these multiple comparisons such as Bonferroni adjustments are not applicable in our case. This is particularly true for the corresponding universal null hypothesis that no statistical relationship exists between flood events and epidemic outbreaks for any of the $\Delta T$ \cite{rothman1990no,perneger1998s}. In contrast, the two detected clusters of $\Delta T$ with statistically significant rates for both precursor and trigger coincidences around monthly and annual time scales indicate the existence of robust coincidence relationships that are present in the data. To further investigate the robustness of these findings, we performed Monte Carlo simulations to assess the probability that the null hypothesis is falsely rejected for fixed pairs of Poisson data surrogates for $n$ values of $\Delta T$. We find that the probability to observe $n=4$ falsely rejected tests at a significance level of 99\,\% (compare Fig.~\ref{fig:results_epidemics_test}) in this setting is less than 0.001, implying that the results presented in Fig.~\ref{fig:results_epidemics_test} can be considered highly statistically significant when taking the effects of multiple testing in the specific setting of our study into account.

\begin{figure}[tb]
	\centering
	\includegraphics[width=0.75\textwidth]{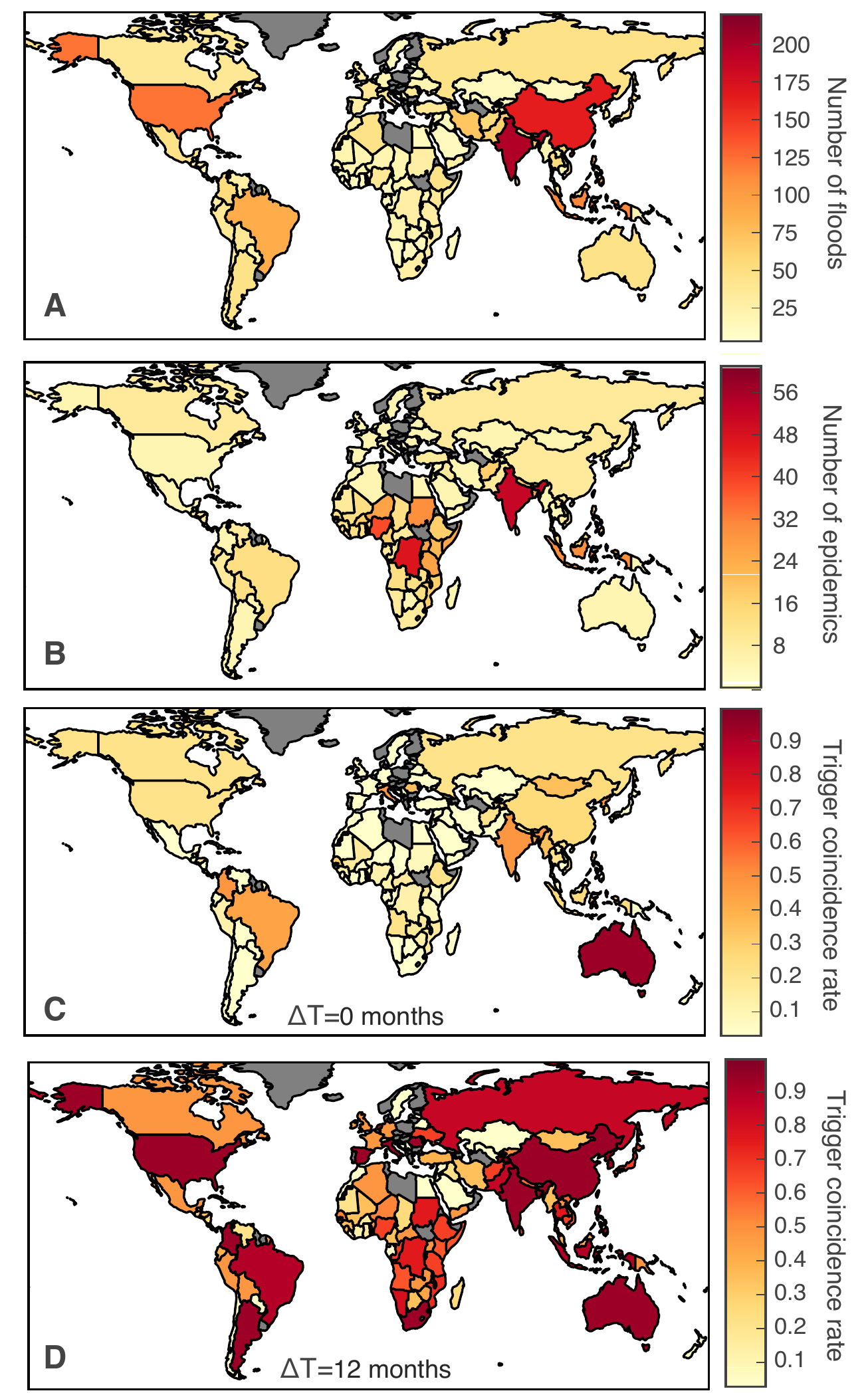}
	\caption{Global mapping of the frequency of floods (A) and epidemics (B) between 1950 and 2009. Country-wise trigger coincidence rates are shown for coincidences occurring within the same month (C) as well as a coincidence interval of 12 months (D). Gray fillings indicate a lack of data for the corresponding countries.}
	\label{fig:map_disaster}
\end{figure}

The global frequencies of flood events and epidemics are depicted in Fig.~\ref{fig:map_disaster}A,B, and the country-wise trigger coincidence rates for coincidences within the same month and a 12-months coincidence interval in Fig.~\ref{fig:map_disaster}C,D. While these maps give some guidance on where floods may have triggered epidemic outbreaks, they need to be interpreted with great caution, since no information about the statistical significance of these rates is conveyed. Since coincidence rates are plotted, countries with very limited statistics (e.g. only one or very few events) can still exhibit high individual coincidence rates.
However, several epidemic-prone regions such as parts of South America, South-East Asia, India and Sub-Saharan Africa are highlighted as having substantial trigger coincidence rates for both coincidence intervals. At the same time, these maps also illustrate a limitation of the tests performed here, since the data is provided on a country-wise resolution, while the considered events, in particularly floods, are bound to geographical regions and water-sheds. This is in particular problematic for larger countries, where a sub-country resolution would be needed to ensure at least the possibility of a causal relation between the event time series. While this represents a clear limitation, it does not affect the significance of our results. The reason is that inclusion of causally unrelated events on a country basis can only increase the probability of coincidences occurring by chance, thereby increasing the significance levels and rendering the test more conservative.

While being the most common natural disasters, floods are the leading cause of natural disaster fatalities worldwide: Doocy \textit{et~al.}~\cite{Doocy2013a} estimate global fatalities due to flood events directly to exceed half a million for the period 1980--2009. At the same time, flood events are also found to increase the risk of outbreaks of fecal-oral, vector-borne and rodent-borne diseases \cite{Ahern2005a}. However, the interrelation between floods and disease outbreaks is found to be complex and strongly case-dependent \cite{Ahern2005a} and, as a consequence, difficult to assess in an aggregated fashion using classical statistical methods. The event-based event coincidence analysis applied here provides a methodological alternative by assessing the statistical interrelationships between the two types of event time series on a case-to-case basis. 
In line with a systematic review of the literature on floods and human health \cite{Alderman2012}, we report robust evidence for both short-term and long-term impacts of floods on epidemic outbreaks. Specifically, we find that more than 50\,\% (20\,\%) of all epidemic outbreaks have been preceded by a flooding event within a 12(1)-month(s) window before the outbreak and that about 20\,\% (7\,\%) of all floods might have triggered such an outbreak in the 12(1) month(s) following the natural disaster. Our results indicate statistically significant coincidence rates up to three months following the disaster and then between 9 and 12 months afterwards, indicating the importance of seasonal effects, which shall be further studied in future work. In particular in tropical regions, flooding events are tied to the rainy season as are major drivers in particular vector-borne diseases~\cite{watson2007epidemics}. Thus, while the significant short-term coincidence rates might to a large extent be a direct consequence of the flooding events, indirect effects will likely dominate the long-term coincidence rates observed, e.g. through impacts on general health, food systems and livelihoods exacerbating poverty and potentially malnutrition that increase long-term susceptibility for diseases  \cite{Alderman2012}. It is important to note, however, that the clustering of floods and epidemics during the rainy season in tropical countries could lead to statistically significant long-term coincidence rates due to the counting of causally unrelated events in successive rainy seasons. This effect should be controlled for in future studies.

Given the projected increase in flood risk under anthropogenic climate change \cite{Arnell2014d,Hirabayashi2013}, our findings highlight the risk of such natural disasters for human health and call for an integrated view on climate and health risks in adaptation efforts. As a note of caution, we would like to stress again the illustrative nature of the results of event coincidence analysis presented for this particular application. More detailed analyses including additional and independent data bases and taking into account systematic effects such as biases induced by changes in self-reporting behavior as have been reported for  EmDAT and other data bases~\cite{saulnier2015systematic} are relevant subjects of future research.

%
%
\section{Conclusions}
\label{sec:conclusions}

In this work, we have introduced event coincidence analysis as a method for investigating statistical interrelationships between event time series such as climate extremes, natural disasters or civil conflicts and other sources of event-like data. Event coincidence analysis builds upon already established methodologies such as event synchronization or measures of correlation between spatial point processes and allows to quantify the strength (via the coincidence rate), directionality (by distinguishing precursor and trigger coincidences) and lag of such interrelationships. Statistical significance tests for these properties have been proposed based on different kinds of null hypotheses on the nature of the temporal point processes underlying the event series, including Poisson processes and stochastic point processes with a given inter-event time distribution. 

As an exemplary application in the timely context of global anthropogenic climate change, we have employed event coincidence analysis for studying statistical interrelationships between  flood events and epidemic outbreaks in the same country on a globally aggregated level. We have found evidence that flood events may have acted as possible drivers of epidemic outbreaks in the past, underlining this potential causal relationship as an important subject of further studies in climate impact and adaptation research.

Promising further methodological developments include the design and more detailed mathematical analysis of appropriate null hypotheses for event coincidence analysis including analytical derivations of the corresponding test statistics as well as the incorporation of event amplitude information~\cite{Suzuki2010}, i.e. by considering marked point processes. Spatial information could be taken into account more explicitly than is the case for the aggregated coincidence rates studied in this paper, building upon a notion of spatio-temporal coincidences with links to the theory of spatial~\cite{Moller2003} and spatio-temporal point processes~\cite{diggle2006spatio}. Furthermore, multivariate extensions such as partial or conditional event coincidence analysis~\cite{Siegmund2016} for measuring statistical interrelations between two event series conditional on a third or even more event series, e.g. methods extending upon the PC-algorithm and its variants~\cite{Runge2015}, would allow to extract further information from rich sources of event data in hypothesis-driven as well as exploratory research modes~\cite{Brown2004}.

%
%
\begin{acknowledgement}
This research was performed in the context of flagship project COPAN on Coevolutionary Pathways in the Earth system and the BMBF Young Investigators Group ``Complex Systems Approaches to Understanding Causes and Consequences of Past, Present and Future Climate Change'' at the Potsdam Institute for Climate Impact Research. We appreciate funding by a Humboldt University / IRI THESys fellowship, the Stordalen Foundation (via the Planetary Boundary Research Network PB.net), the Earth League's EarthDoc program, the German Federal Ministry for Education and Research (BMBF projects GLUES and CoSy-CC$^2$ (grant no. 01LN1306A)) and the Evangelisches Studienwerk Villigst. The work was supported by the German Federal Ministry for the Environment, Nature Conservation and Nuclear Safety (11-II-093-Global-A SIDS and LDCs). Jobst Heitzig, Marc Wiedermann and Miguel Mahecha are acknowledged for helpful insights and discussions at various stages of the reported research. Event coincidence analyses can be performed using the R package \texttt{CoinCalc} \cite{Siegmund2015b} which is available at \texttt{https://github.com/JonatanSiegmund/CoinCalc}.
\end{acknowledgement}

%

%
%

\bibliographystyle{epj}
\bibliography{coincidences}

\end{document}